# Beam-Size Invariant Spectropolarimeters Using Gap-Plasmon Metasurfaces

Fei Ding,[§,*] Anders Pors,[§,*] Yiting Chen, Vladimir A. Zenin, and Sergey I. Bozhevolnyi

Centre for Nano Optics, University of Southern Denmark, Campusvej 55, DK-5230 Odense, Denmark

[§]These authors contributed equally to this work.

[*]To whom correspondence should be addressed. Email address: feid@iti.sdu.dk.

**ABSTRACT**

Metasurfaces enable exceptional control over the light with surface-confined planar components, offering the fascinating possibility of very dense integration and miniaturization in photonics. Here, we design, fabricate and experimentally demonstrate chip-size plasmonic spectropolarimeters for simultaneous polarization state and wavelength determination. Spectropolarimeters, consisting of three gap-plasmon phase-gradient metasurfaces that occupy 120° circular sectors each, diffract normally incident light to six predesigned directions, whose azimuthal angles are proportional to the light wavelength, while contrasts in the corresponding diffraction intensities provide a direct measure of the incident polarization state through retrieval of the associated Stokes parameters. The proof-of-concept 96-μm-diameter spectropolarimeter operating in the wavelength range of 750 – 950 nm exhibits the expected polarization selectivity and high angular dispersion (0.0133 °/nm). Moreover, we show that, due to the circular-sector design, polarization analysis can be conducted for optical beams of different diameters without prior calibration, demonstrating thereby the beam-size invariant functionality. The proposed spectropolarimeters are compact, cost-effective, robust, and promise high-performance real-time polarization and spectral measurements.

**KEYWORDS**

Spectropolarimeters, metasurface, beam-size invariant, gap-plasmon

**Table of Content**

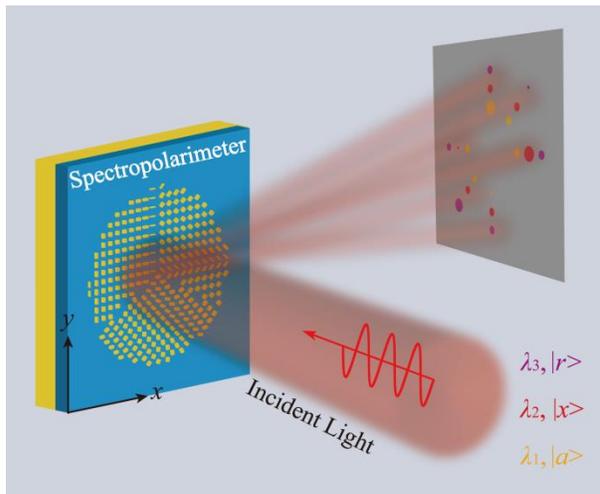

Light is an important tool to instigate significant technological advancements through analysis of its intensity, frequency, and the state of polarization (SOP). Spectropolarimeters, which enable simultaneous measurements of the spectrum and SOP of light, constitute a powerful analytic tool far exceeding capabilities of separate polarimeters and spectrometers with their information channels being uncorrelated. Spectropolarimeters find important applications in many areas of science and technology, ranging from astronomy[1] and medical diagnostics[2] to remote sensing,[3] since the SOP carries rich information about the composition and structure of materials interrogated with light that is beyond the capability of spectroscopic (or polarimetric) analysis.

Despite all scientific and technological potential, spectropolarimetry is still very challenging to experimentally realize as the SOP characterization alone requires conventionally six intensity measurements to determine the Stokes parameters.[4] This problem is typically addressed by sequential measurements combining spectrometers, which involve dispersive elements (gratings or prisms), with rotating retarder polarimeters. As such, the SOP is commonly evaluated utilizing a set of properly arranged polarization elements, for example, polarizers and waveplates, which are consecutively placed in the light path in front of a detector. In this way, the Stokes parameters that uniquely define the SOP are determined by measuring the light flux transmitting through these polarization components. Consequently, spectropolarimeters based on conventional discrete optical components amount to bulky, complex and expensive optical systems that are not compatible with the general trend of integration and miniaturization in photonics.

In recent years, optical metasurfaces, i.e., designed and fabricated surface nanostructures that modify boundary conditions for impinging optical waves in order to realize specific wave transformations, have been gaining increasing attention due to their remarkable abilities in light manipulation, their versatility, ease of on-chip fabrication and integration owing to their planar profiles.[5,6] Such metasurfaces can mimic bulk optics since they are capable of engineering the phase front of reflected and/or refracted optical waves at will. Many ultra-compact flat optical components have been accordingly demonstrated, such as beam steerers,[7-10] surface waves couplers,[11-13] focusing lenses,[14-16] optical holograms,[17-20] waveplates,[21-24] and spectrometers.[25]

Metasurfaces represent therefore an opportunity for spectropolarimetry to overcome bulky and expensive architectures based on conventional volume optics. In early approaches, metasurfaces together with conventional optical elements, such as polarizers and retardation waveplates,[26] or the effect of a polarization-dependent transmission of light through six carefully designed nano-apertures in metal films,[27] were designed for the purposes of polarimetry. Additionally, different types of metasurfaces were proposed to determine certain aspects of the SOP, like the degree of circular polarization.[28-29] Recently, on-chip metasurface-only polarimeters allowing for the simultaneous determination of all polarization states have been proposed and demonstrated,[30-32] but the implemented approaches are not suitable for extending to also conduct spectral analysis. Related developments have revealed that segmented[33] and interleaved[34] metasurfaces can be designed to conduct spectropolarimetry with simultaneous characterization of the SOP and spectrum of optical waves. But these configurations require careful calibrations. Additionally, segmented rectangular configuration[33] are oriented towards the spectropolarimetry of plane waves, so that the incident light should cover the whole area of metasurfaces with the same light intensity over all metasurface elements in order to ensure faithful comparison of the corresponding (to particular Stokes parameters) diffraction orders.

Here, we design, fabricate and experimentally demonstrate chip-size plasmonic spectropolarimeters for simultaneous SOP and spectral detection consisting of three gap-plasmon phase-gradient metasurfaces that occupy 120° circular sectors each. The main idea is that this center-symmetrical configuration would diffract any normally incident (and centered) beam with a circular cross-section to six predesigned directions, whose azimuthal angles are proportional to the light wavelength, while contrasts in the corresponding diffraction intensities would provide a direct measure of the incident polarization state through retrieval of the associated Stokes parameters. The fabricated spectropolarimeters operating in the wavelength range of 750 – 950 nm are found to exhibit the expected polarization sensitivity, which is also beam-size invariant, and, in the case of the 96-µm-diameter metasurface, reach the angular dispersion of 0.0133 °/nm.

**RESULTS AND DISCUSSION**

The working principle of the proposed spectropolarimeters that should selectively steer the different input SOP and spectral components into six separate spatial domains is schematically illustrated in Figure 1a. In contrast to the previous interweaved metagrating[30] and rectangular metasurface array,[33] three different types of phase-gradient birefringent metasurfaces occupy 120° circular sectors each, being thereby incorporated in a large circular configuration. Each sector-shaped metasurface functions as an efficient polarization splitter for one of the three different polarization bases [$(\vec{x}, \vec{y})$, $(\vec{a}, \vec{b})$ and $(\vec{r}, \vec{l})$]. The basis $(\vec{a}, \vec{b})$ corresponds to a rotation of the Cartesian coordinate system $(\vec{x}, \vec{y})$ by 45° with respect to the $x$ axis, while $(\vec{r}, \vec{l})$ is the basis for circularly polarized light. Based on the pronounced diffraction contrast in the $(\vec{u}, \vec{v})$ basis, defined by $D = \frac{I_{+1} - I_{-1}}{I_{+1} + I_{-1}} = \frac{A_u^2 - A_v^2}{A_u^2 + A_v^2}$, where $I_{+1} \propto A_u^2$ and $I_{-1} \propto A_v^2$ are the corresponding intensities in the ±1 diffraction orders, and $(\vec{u}, \vec{v})$ represents one of the three bases $(\vec{x}, \vec{y})$, $(\vec{a}, \vec{b})$ and $(\vec{r}, \vec{l})$, one can readily retrieve the corresponding Stokes parameters[30] that fully describe the SOP of interrogated optical wave (see Supplementary Part 1 for details). Furthermore, since the diffraction angle is determined by the light wavelength (and by the supercell size[10,30]), the azimuthal angle of diffraction orders can be used to determine the light wavelength, thus gaining the spectroscopic information.

To design the gap-plasmon based phase-gradient metasurface, we first consider the classical metal-insulator-metal configuration without phase gradient shown in Figure 1b. The unit cell is composed of gold (Au) nanobricks arranged periodically in the $x$-$y$ plane and a continuous Au film, separated by a silicon dioxide ($SiO_2$) dielectric layer. Here the nanobrick dimensions ($L_x$ and $L_y$) represent the two variable parameters for phase-gradient design, whereas the other geometrical dimensions are constant. The other parameters of the unit cell are: $\Lambda_x$ = 320 nm, $\Lambda_y$ = 250 nm, $d$ = 120 nm, $t_s$ = 50 nm and $t$ = 40 nm. We implement three-dimensional (3D) full wave simulations with the commercially available software Comsol Multiphysics (ver. 5.2) to compute the reflection coefficient for both the *x*- and *y*-polarized light. In the simulation, the permittivity of Au is described by interpolated experimental values,[35] whereas the $SiO_2$ spacer is taken as a lossless dielectric with a constant refractive index of 1.45. Periodic boundary conditions are used in the *x*- and *y*-directions, and a plane wave is incident downward on the structure with the electric field polarized along the *x*-direction (or *y*-direction) as the excitation source.

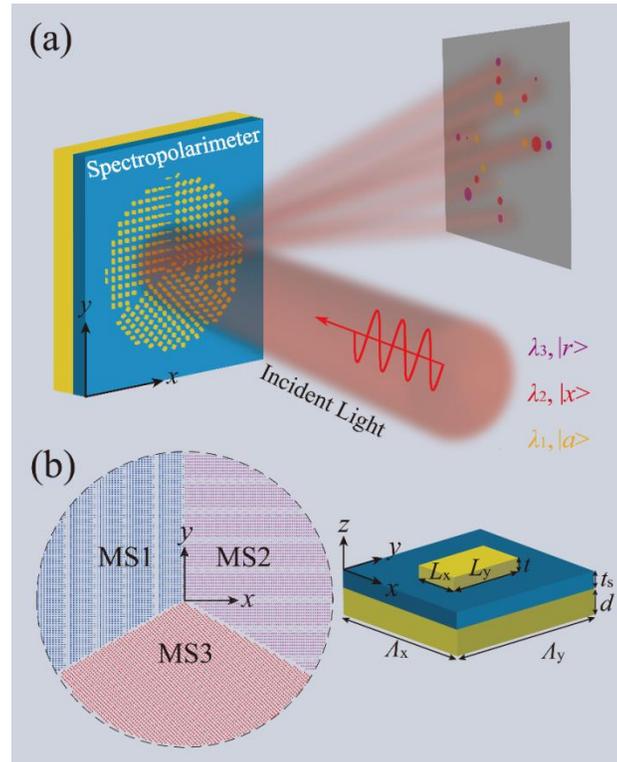

**Figure 1.** Illustration of the gap-plasmon metasurface based spectropolarimeter. (a) Artistic rendering of the working principle: different polarization and spectral components are selectively diffracted into pre-designed spatial domains with distinct spot contrasts. (b) Top view of the spectropolarimeter composed of three phase-gradient metasurfaces shown in different colors. The right panel shows the schematic of the unit cell consisting of a gold nanobrick on top of a spacer and gold substrate. The fixed geometrical parameters are $\Lambda_x$ = 320 nm, $\Lambda_y$ = 250 nm, $d$ = 120 nm, $t_s$ = 50 nm, and $t$ = 40 nm.

In the design of metasurface 1 (MS1) that splits the pair of orthogonal polarization states in $(\vec{x}, \vec{y})$ basis, we discretize the $2\pi$ phase range into 10 equal steps and the lateral dimensions of the selected 10 nanobricks are shown in Figure 2a. In Figure 2b the black and blue curves correspond to the co-polarized reflection amplitudes and phases of the 10 nanobricks for the linearly polarized incident light. One can clearly see that the supercell consisting of 10 nanobricks provides an incremental (decremental) phase of $\pi/5$ from Element 1 to Element 10 for the co-polarized reflected light when the incident light is *x*-polarized (*y*-polarized), resulting in anomalous diffraction into angles of $\pm 14.5°$, respectively. It should be noted that although the linear phase-gradient is designed for a nominal wavelength at 800 nm, it exhibits only weak wavelength-dependence, allowing broadband beam-splitting.[7,10] In addition, the non-ideal reflection amplitude distributions have little effect on the performance. The second metasurface (MS2), intended to split light in the basis $(\vec{a}, \vec{b})$, can be designed in a completely equivalent way as described for MS1, except that each nanobrick element is rotated counterclockwise by 45° with respect to the *x*-axis (Figure 2c). Finally, the third metasurface (MS3), which selectively separates the circularly polarized light with opposite helicity, is constructed by implementing the geometric phase covering all $2\pi$ phase range with 10 different orientations shown in Figure 2c. To increase the efficiency of MS3, the nanobrick unit cell (Element 11 in Figure 2a) is optimized as highly-efficient half-wave plate.[20,24] In order to validate the polarization-sensitive diffraction of these three metasurfaces, we have conducted full-wave simulations to calculate the corresponding diffraction efficiencies (see Supplementary Part 2). As expected, the first-order diffraction contrasts remain high in the wavelength range of 750-950 nm for all three cases, thereby implying that the designed metasurfaces can operate in a broad wavelength range.

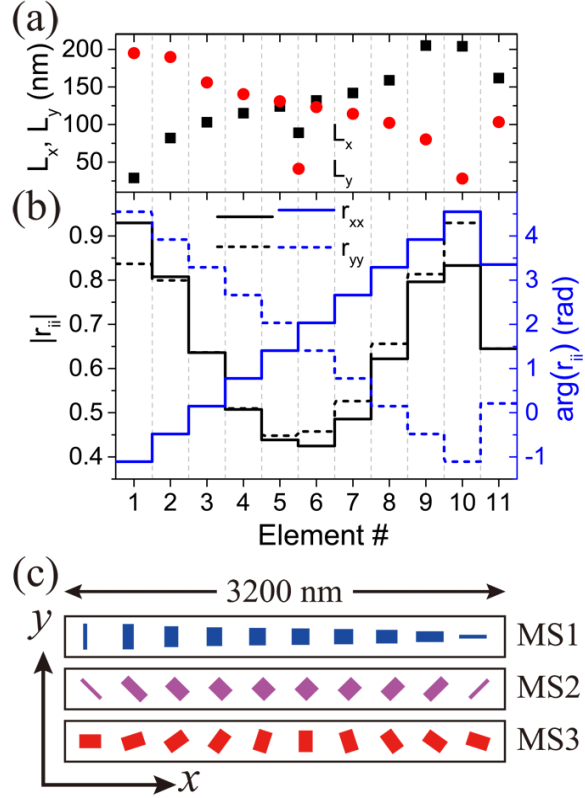

**Figure 2.** Design of the gap-plasmon metasurfaces. (a) Lateral dimensions of the 11 different nanobrick elements used in the design. The first 10 nanobricks (Element 1 to Element 10) are designed to produce equidistant (within $2\pi$ phase range) phase shift. Element 11 is the designed half-wave plate for the circularly polarized light. (b) The corresponding reflection amplitudes and phases of the associated 11 nanobricks at $\lambda = 800$ nm. (c) The three metasurface supercells constituted by the 11 different nanobricks for the polarization bases $(\vec{x}, \vec{y})$, $(\vec{a}, \vec{b})$ and $(\vec{r}, \vec{l})$.

To experimentally validate the simultaneous SOP and spectral analysis, several spectropolarimeters of different sizes were fabricated using standard electron beam lithography (EBL) and a lift-off process (see more details of fabrication in Methods section). Figure 3a displays a scanning electron microscope (SEM) image of the largest fabricated configuration with the diameter of 96 $\mu$m. Although smaller diameters imply smaller device footprints, one should bear in mind that the achievable spectral resolution is proportional to the size of a metasurface, with the resolving power being roughly equal to the number of supercells across the metasurface (just like for diffraction gratings).[34,36]

Following fabrication, we characterized the spectropolarimeter using the home-made setup which captures the far-field intensities of the predetermined polarization states (details of the far-field optical characterization are given in Methods section and Supplementary Part 3). It should be noted that the overall uncertainty in the SOP of the incident light beam is estimated to be nearly 10%, which is ascribed to the imperfect performance of the optical components as well as non-ideal optical alignment.

Figure 3c shows the relevant far-field images of the diffraction spots for six different incident SOPs, each resulting in a unique intensity distribution. We emphasize that, distinct from related work,[30,37] there is no unwanted diffraction in the $yz$ plane, produced by interweaving three metasurfaces into one single meta-grating. The noticeable zero-order diffraction may be ascribed to the imperfections and surface roughness of the fabricated nanobricks (Figure 3a), the gap between adjacent metasurfaces, together with the uncertainty of practical optical constant of evaporated Au film. From the CCD images, one can clearly see that the bright spots from the ±1 diffraction order change in intensity in response to altering the incident SOP. For instance, the spot intensity of channel 1 is gradually suppressed, whereas the spot of channel 4 becomes brighter and brighter as the polarization angle of the incident linearly polarized light is continuously varied from $|x\rangle$ state to $|y\rangle$ state through $|a\rangle$ and $|b\rangle$ states (see the Supplementary Movie 1 for details). When the incident circularly polarized light is switched from $|r\rangle$ to $|l\rangle$, channel 5 becomes dominated while channel 2 is nearly dark. Based on these experimental images, the calculated diffraction contracts are shown in Figure 3d. In general, good agreement is observed between the measured diffraction contrast and the normalized Stokes parameters for linear polarization states, considering the aforementioned fabrication imperfections, uncertainties in the material properties, as well as the uncertainty of the input SOP. However, when the incident light becomes circularly polarized, the discrepancy increases a little bit, which is mainly attribute to the non-ideal input polarization state affected by the imperfect performance of the optical components, particularly the broadband quarter-wave plate, together with inevitable alignment and rotation errors. Thus, input light is elliptically polarized rather than circularly polarized, increasing the discrepancy. More experimental results with different input SOPs are presented in Supplementary Part 4, validating the polarization selectivity. We would like to emphasize that the overall accuracy of SOP determination can be significantly improved, if one would adopt the procedure based on careful device calibration[33,34] instead of simply using the normalized diffraction contrasts. The calibration procedure is well known[38] and can always be employed once the polarization selectivity is established.[33,34]

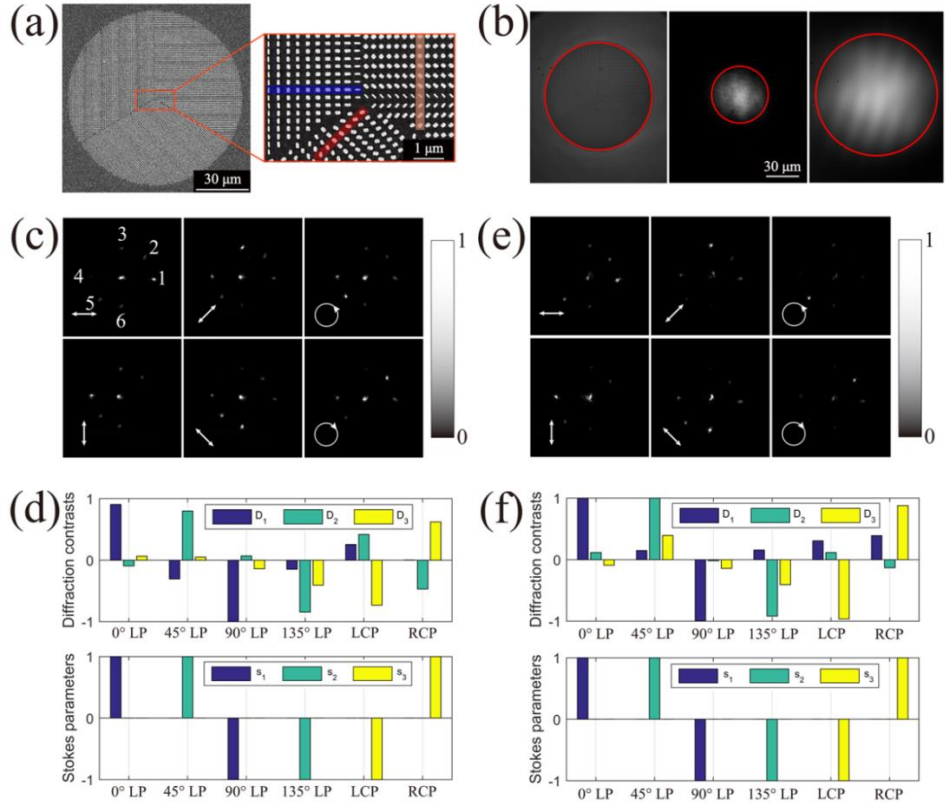

**Figure 3.** Experimental implementation of the metasurface-based spectrapolarimeter. (a) SEM image of the fabricated sample (scale bar 30 μm). Local magnification of the central area is also shown (scale bar 1 μm). (b) CCD images of the sample under the illumination of white light (left) and titanium sapphire laser with different beam sizes (middle and right) respectively. All the images have the same scale (scale bar 30 μm). (c and e) Normalized far-field images for the six incident SOPs at $\lambda$ = 800 nm with the incident beam spot sizes of 30.7±1.7 μm and 71 ± 4.6 μm, respectively. The incident SOPs are shown as white arrows and correspond to the six extreme polarizations. The brightness and contrast have been adjusted for visualization. The six output channels of diffraction spots are marked by white numbers in (c). (d and f) Measured diffraction contrasts for the six polarizations that represent the extreme value of the three Stokes parameters with the incident beam spot sizes of 30.7±1.7 μm and 71 ± 4.6 μm, respectively.

Unlike previous demonstration with six metasurfaces arranged in a 2 × 3 array,[33] the performance of our spectropolarimeter is robust with respect to the spot size of the incident laser beam, i.e., it is beam-size invariant, at least with respect to the SOP determination, a feature which is ascribed to the particular and meticulous design. Our spectropolarimeter is realized by incorporating three metasurfaces occupying 120° circular sectors each into one large circular configuration, with each sectorial metasurface being subjected to the same incident power provided that the incident beam is centered and has the (polar) angle-independent intensity distribution. Therefore each metasurface, operating independently as a polarization splitter for one of the polarization pairs ($|x\rangle$, $|y\rangle$), ($|a\rangle$, $|b\rangle$) and ($|r\rangle$, $|l\rangle$), diffracts the same power allowing for a direct use of the diffraction contrasts as measures of the Stokes parameters. In order to validate the beam-size invariant property, we increase the beam spot size to 71±4.6 μm [right panel in Figure 3b], which is about twice of the previous size of 30.7±1.7 μm [middle panel in Figure 3b]. The corresponding far-field images and the measured diffraction contrasts are shown in Figure 3e and 3f, respectively. As expected, the intensity distributions of the six characteristic diffraction spots remain the same while the unwanted zeroth diffraction order becomes slightly weak. After integrating the far-field intensities, we find that approximately ~60% of the refracted light go into the relevant ±1 diffraction order, which is a little higher than the value of ~45% when the incident beam spot size is 30.7±1.7 μm. Considering the total reflection efficiency of our device (~80% in simulation), the total efficiency is found to be at most equal to ~48%. Moreover, the measured diffraction contrasts stay closer to the normalized Stokes parameters when the spot size is doubled, resulting from the more homogeneous incident laser beam. Therefore, with respect to the beam size and non-uniform intensity distribution, our design exhibits advantageous features in terms of robustness and stability as the functionality doesn't critically depend on the size and inhomogeneity of the laser beam.

As a spectropolarimeter, simultaneous measurement of the spectrum and polarization state of light is needed. To validate the capability of spectral analysis besides polarization probe, we implemented more measurements within a wavelength range from 750 nm to 950 nm. Figure 4a-b present the associated diffraction contrasts for the six polarizations that represent the three normalized Stokes parameters at $\lambda$ = 750 nm and 850 nm, respectively. It can be clearly observed that the diffraction contrasts are positioned close to the corresponding normalized Stokes parameters when the input light is linearly polarized, demonstrating good polarization analysis in a broad spectral range. In addition, the performance of circular polarization probe is slightly depressed because of the increasing uncertainty in the state of circular polarization related with the quarter-wave plate.

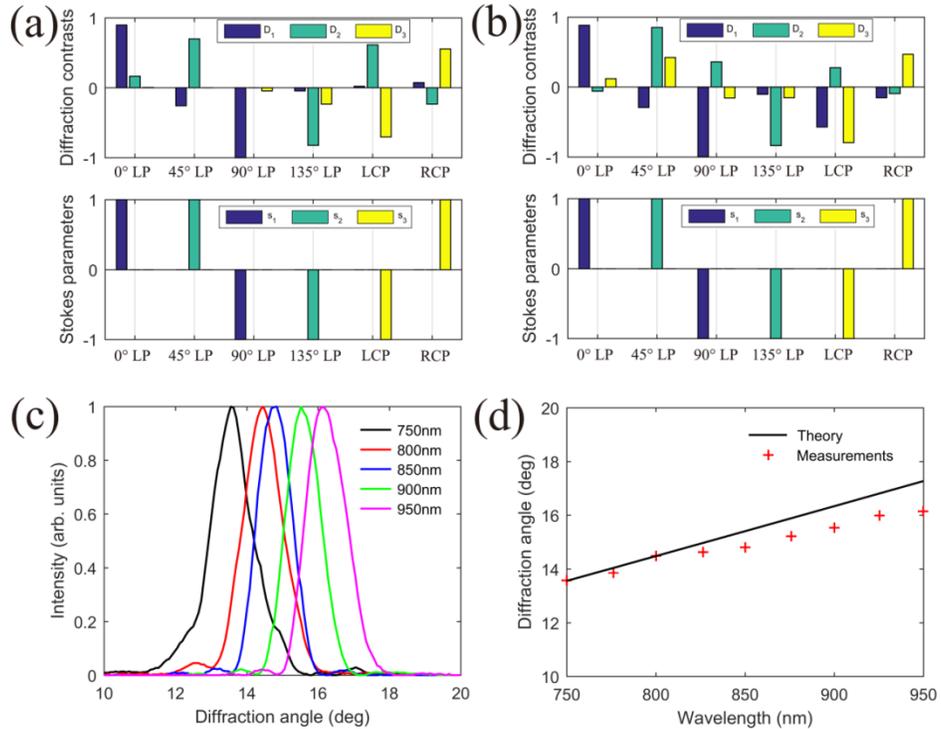

**Figure 4.** (a and b) Measured diffraction contrasts for the six polarizations that represent the extreme value of the three Stokes parameters at $\lambda$ = 750 nm and 850 nm, respectively. (c) Normalized measured far-field intensity profile for different wavelength of $|x\rangle$ channel. (d) Theoretical predictions and experimentally measured spectral dispersion for $|x\rangle$ channel.

We subsequently studied the spectral resolution of our device, which is of significant importance for the spectroscopic measurements, including spectropolarimetry. In our design, the spectral resolution is closely related to the angular dispersion $\Delta\theta/\Delta\lambda$. Since the spectropolarimeter shows good polarization and spectral analysis for linear polarization state, we, therefore, take the $|x\rangle$ channel (channel 1) as an example to analyze the angular dispersion. From Figure 4c, we note that the polarization selectivity is maintained over a 200 nm-wide bandwidth. The bright characteristic spot moves outwards once the incident wavelength is increased (see the Supplementary Movie 2 for switching wavelengths). Figure 4d suggests that the experimentally measured angular dispersion $\Delta\theta/\Delta\lambda$ for the $|x\rangle$ channel is 0.0133 °/nm, which is in good agreement with the theoretical prediction (black solid curve). If this device is used as the dispersive element in a spectrometer with the same assumption of Ref. 33, the estimated spectral resolution could be ~1.08 nm if the dispersive device is infinite. However, this assumption cannot precisely predict the spectral resolution power as the diameter of our device is only 96 μm. Specifically, the spectral resolution power can be defined as $\frac{\lambda}{\Delta\lambda} = \frac{q}{M_{x,y}^2}$ in the case of a finite-size dispersive device, where $q$ is the number of phase modulation periods and $M_{x,y}^2$ is the beam quality parameter.[34,36] The measured spectral resolving power here is about 15.2 if we take 0.7° as the actual minimum resolvable angular difference from the Rayleigh criterion, which is in good agreement with the calculated resolving power with $q$ = 15 periods and $M_{x,y}^2 = 1$. Additionally, the resolving power is comparable to that of Ref. 34. It's worth noting that the spectral resolving power can be further improved with proper design, for instance, designing interleaved metasurface instead of segmented sample and using super-dispersive off-axis meta-lenses that simultaneously focus and disperse light of different wavelengths,[25] even though the size of spectropolarimeter is fixed.

**CONCLUSIONS**

In conclusion, we have proposed and demonstrated chip-size ultra-compact plasmonic spectropolarimeters consisting of three gap-plasmon phase-gradient metasurfaces that occupy 120° circular sectors each, allowing the fast and simultaneous SOP and spectral determination in one measurement. The proof-of-concept 96-μm-diameter spectropolarimeter operating in the wavelength range of 750 – 950 nm exhibits the expected polarization selectivity with high angular dispersion (0.0133 °/nm). Importantly, due to the circular-sector design, polarization analysis can be implemented for optical beams of different diameters without prior calibration, demonstrating thereby the beam-size invariant functionality. Our proposed spectropolarimeters can furthermore be scaled to operate at other relevant frequencies, such as the technologically difficult terahertz frequency. Owing to the compactness, integration compatibility and robustness, our spectropolarimeters promise high performance real-time polarization and spectrum monitoring for more advanced applications, for instance, drug manufacturing, medical imaging, quantum communication and astronomy.

**METHODS**

**Fabrication of spectropolarimeter.** First, the successive layers of 3 nm Ti, 120 nm Au, 3 nm Ti and 50 nm $SiO_2$ are deposited onto silicon substrate using electron-beam evaporation (metals) and RF-sputtering ($SiO_2$). Then, the sample was coated with a 100-nm-thick e-beam resist PMMA (2% in anisole, Micro Chem) layer. Subsequently, the metasurface was defined using e-beam lithography at the acceleration voltage of 30 keV. After exposure, the sample was developed in solution of methyl isobutyl ketone (MIBK) and isopropyl alcohol (IPA) of MIBK: IPA=1:3 for 35 seconds. Once the development of the resist was complete, a 3 nm Ti adhesion layer and a 40 nm gold layer are deposited subsequently using electron beam deposition. After lift-off process was performed using acetone, the Au patterns are finally formed on top of the SiO2 film.

**Far-field Optical Characterization.** The sample is mounted on a stage with XYZ translation and exposed to the laser beam from a tunable Ti: Sapphire laser with wavelength set to be 800 nm. At the same time, the sample is illuminated with white light for visualization (see Figure S2 for details). The polarization state of the incident light is controlled by two polarizers (P1 and P2) and wave plates (half-wave or quarter-wave plate, WP). Once the polarization state is fixed, the light is weakly focused by a lens (L1) onto the sample with a spot smaller than the metasurface. The reflected light is collected by a long working distance objective, whose numerical aperture (NA) is 0.55. The front focal plane is located at the surface of the sample. The diffusion property of the metasurface is finally obtained by projecting the objective back focal plane (BFP) of the objective by another lens (L2) onto a CMOS camera with high sensitivity. In order to compensate the phase change induced by BS2 between the two orthogonally polarizations, an additional beam splitter (BS1), which is rotated by 90° with respect to BS2, is inserted in the optical path, preserving the polarization state of the incident light.

**Acknowledgements**

This work was funded by the European Research Council (the PLAQNAP project, Grant 341054) and the University of Southern Denmark (SDU2020 funding).

**Competing financial interests:**
The authors declare no potential conflicts of interest.

**Supporting Information**

The Stokes parameters, the theoretical performance of the phase-gradient metasurfaces, optical setup of far-field characterization, measured diffraction contrasts for other polarization states, and evaluation of the incident beam size. This material is available free of charge *via* the Internet at http://pubs.acs.org.

**References**

(1) Sterzik, M. F.; Bagnulo, S.; Palle, E. Biosignatures as revealed by spectropolarimetry of Earthshine. *Nature* **2012**, *483*, 64-66.

(2) Gupta, N. J. Biomed. Acousto-optic-tunable-filter-based spectropolarimetric imagers for medical diagnostic applications—instrument design point of view. *J. Biomed. Opt.* **2005**, *10*, 051802-051802-6.

(3) Tyo, J. S.; Goldstein, D. L.; Chenault, D. B.; Shaw, J. A. Review of passive imaging polarimetry for remote sensing applications. *Appl. Opt.* **2006**, *45*, 5453-5469.

(4) Bohren, C. F.; Huffman, D. R. *Absorption and Scattering of Light by Small Particles*; Wiley: New York, 1983.

(5) Kildishev, A. V.; Boltasseva, A.; Shalaev, V. M. Planar Photonics with Metasurfaces. *Science* **2013**, *339*, 1232009.

(6) Yu, N.; Capasso, F. Flat optics with designer metasurfaces. *Nat. Mater.* **2014**, *13*, 139–150.

(7) Yu, N.; Genevet, P.; Kats, M. A.; Aieta, F.; Tetienne, J. P.; Capasso, F.; Gaburro, Z. Light Propagation with Phase Discontinuities: Generalized Laws of Reflection and Refraction. *Science* **2011**, *334*, 333−337.

(8) Ni, X.; Emani, N. K.; Kildishev, A. V.; Boltasseva, A.; Shalaev, V. M. Broadband Light Bending with Plasmonic Nanoantennas. *Science* **2012**, *335*, 427−427.

(9) Pfeiffer, C.; Grbic, A. Metamaterial Huygens' Surfaces: Tailoring Wave Fronts with Reflectionless Sheets. *Phys. Rev. Lett.* **2013**, *110*, 197401.

(10) Pors, A.; Albrektsen, O.; Radko, I. P.; Bozhevolnyi, S. I. Gap plasmon-based metasurfaces for total control of reflected light. *Sci. Rep.* **2013**, *3*, 2155.

(11) Sun, S.; He, Q.; Xiao, S.; Hu, Q.; Zhou, L. Gradient-index meta-surfaces as a bridge linking propagating waves and surface waves. *Nat. Mater.* **2012**, *11*, 426–431.

(12) Lin, J.; Mueller, J. B.; Wang, Q.; Yuan, G.; Antoniou, N.; Yuan, X. C.; Capasso, F. Polarization-Controlled Tunable Directional Coupling of Surface Plasmon Polaritons. Science **2013**, *340*, 331-334.

(13) Pors, A.; Nielsen, M. G.; Bernardin, T.; Weeber, J.-C.; Bozhevolnyi, S. I. Efficient unidirectional polarization-controlled excitation of surface plasmon polaritons. *Light: Sci. Appl.* **2014,** *3*, e197.

(14) Aieta, F.; Genevet, P.; Kats, M. A.; Yu, N.; Blanchard, R.; Gaburro, Z.; Capasso, F. Aberration-Free Ultrathin Flat Lenses and Axicons at Telecom Wavelengths Based on Plasmonic Metasurfaces. *Nano Lett.* **2012**, *12*, 4932−4936.

(15) Ni, X.; Ishii, S.; Kildishev, A. V.; Shalaev, V. M. Ultra-thin, planar, Babinet-inverted plasmonic metalenses. *Light: Sci. Appl.* **2013**, *2*, e72.

(16) Pors, A.; Nielsen, M. G.; Eriksen, R. L.; Bozhevolnyi, S. I. Broadband Focusing Flat Mirrors Based on Plasmonic Gradient Metasurfaces. Nano Lett. 2013, 13, 829–834.

(17) Ni, X.; Kildishev, A. V.; Shalaev, V. M. Metasurface holograms for visible light. *Nat. Commun.* **2013**, *4*, 2807.

(18) Chen, W. T.; Yang, K. Y.; Wang, C. M.; Huang, Y. W.; Sun, G.; Chiang, I. D.; Liao, C. Y.; Hsu, W. L.; Lin, H. T.; Tsai, D. P. High-Efficiency Broadband Meta-Hologram with Polarization-Controlled Dual Images. *Nano Lett.* **2014**, *14*, 225–230.

(19) Patrice, G.; Federico, C. Holographic optical metasurfaces: a review of current progress. *Rep. Prog. Phys.* **2015,** *78*, 024401.

(20) Zheng, G.; Mühlenbernd, H.; Kenney, M.; Li, G.; Zentgraf, T.; Zhang, S. Metasurface holograms reaching 80% efficiency. *Nat. Nano.* **2015,** *10*, 308-312.

(21) Pors, A.; Nielsen, M. G.; Bozhevolnyi, S. I. Broadband plasmonic half-wave plates in reflection. *Opt. Lett.* **2013**, *38*, 513-515.

(22) Zhao, Y.; Alù, A. Tailoring the dispersion of plasmonic nanorods to realize broadband optical meta-waveplates. *Nano Lett.* **2013**, *13*, 1086-1091.

(23) Grady, N. K.; Heyes, J. E.; Chowdhury, D. R.; Zeng, Y.; Reiten, M. T.; Azad, A. K.; Taylor, A. J.; Dalvit, D. A.; Chen, H. T. Terahertz Metamaterials for Linear Polarization Conversion and Anomalous Refraction. *Science* **2013**, *340*, 1304–1307.

(24) Ding, F.; Wang, Z.; He, S.; Shalaev, V. M.; Kildishev, A. V. Broadband High-Efficiency Half-Wave Plate: A Supercell-Based Plasmonic Metasurface Approach. *ACS Nano* **2015**, *9*, 4111-4119.

(25) Khorasaninejad, M.; Chen, W. T.; Oh, J.; Capasso, F. Super-Dispersive Off-Axis Meta-Lenses for Compact High Resolution Spectroscopy. *Nano Lett.* **2016,** *16*, 3732-3737.

(26) Bomzon, Z.; Biener, G.; Kleiner, V.; Hasman, E. Spatial Fourier-transform polarimetry using space-variant subwavelength metal-stripe polarizers. *Opt. Lett.* **2001,** *26*, 1711-1713.

(27) Afshinmanesh, F.; White, J.; Cai, W.; Brongersma, M. L. Measurement of the polarization state of light using an integrated plasmonic polarimeter. *Nanophotonics*, **2012,** *1*, 125-129.

(28) Shaltout, A.; Liu, J.; Kildishev, A.; Shalaev, V. Photonic spin Hall effect in gap plasmon metasurfaces for on-chip chiroptical spectroscopy. *Optica* **2015,** *2*, 860-863.

(29) Wen, D.; Yue, F.; Kumar, S.; Ma, Y.; Chen, M.; Ren, X.; Kremer, P. E.; Gerardot, B. D.; Taghizadeh, M. R.; Buller, G. S.; Chen, X. Metasurface for characterization of the polarization state of light. *Opt. Express* **2015,** *23*, 10272-10281.

(30)   Pors, A.; Nielsen, M. G.; Bozhevolnyi, S. I. Plasmonic metagratings for simultaneous determination of Stokes parameters. *Optica* **2015,** *2*, 716-723.

(31)   Balthasar Mueller, J. P.; Leosson, K.; Capasso, F. Ultracompact metasurface in-line polarimeter. *Optica* **2016,** *3*, 42-47.

(32)   Pors, A.; Bozhevolnyi, S. I. Waveguide Metacouplers for In-Plane Polarimetry. *Phys. Rev. Applied* **2016,** *5*, 064015.

(33)   Chen, W.; Török, P.; Foreman, M.; Liao, C.; Tsai, W.; Wu, P.; Tsai, D. Integrated plasmonic metasurfaces for spectropolarimetry. *Nanotechnology* **2016,** *27*, 224002.

(34)   Maguid, E.; Yulevich, I.; Veksler, D.; Kleiner, V.; Brongersma, M. L.; Hasman, E. Photonic spin-controlled multifunctional shared-aperture antenna array. *Science* **2016,** *352*, 1202.

(35)   Johnson, P. B.; Christy, R. W. Optical Constants of the Noble Metals. *Phys. Rev. B* **1972,** *6*, (12), 4370-4379.

(36)   Davidson, N.; Khaykovich, L.; Hasman, E. High-resolution spectrometry for diffuse light by use of anamorphic concentration. *Opt. Lett.* **1999**, *24*, 1835–1837.

(37)   Lepetit, T.; Kante, B. Metasurfaces: Simultaneous Stokes parameters. *Nat. Photon.* **2015**, *9*, 709-710.

(38)   Azzam, R. M. A.; Lopez, A. G. Accurate calibration of the four-detector photopolarimeter with imperfect polarizing optical elements. *J. Opt. Soc. Am. A* **1989,** *6*, 1513-1521.